\documentclass[a4paper,fleqn,usenatbib]{mnras}

\usepackage{newtxtext,newtxmath}

\usepackage[T1]{fontenc}
\usepackage{ae,aecompl}

\usepackage{graphicx}	
\usepackage{amsmath}	
\usepackage{amssymb}	
\usepackage{booktabs}
\usepackage{threeparttable}
\usepackage[normalem]{ulem}            
\usepackage{xcolor}           

\newcommand{\ha}{H$\alpha$}
\newcommand{\hei}{\ion{He}{i}}
\newcommand{\kms}{km s$^{-1}$}
\newcommand{\macc}{\ensuremath{\dot{M}_{\rm acc}}}
\newcommand{\msun}{M$_{\odot}$}

\newcommand{\vten}{\ensuremath{v_{10}}}
\title[2M1546: a mid-M dwarf hosting a prolonged accretion disc]{2MASS
J15460752-6258042: a mid-M dwarf hosting a prolonged accretion disc}

\author[Lee et al.]{
Jinhee Lee,$^{1,2}$
Inseok Song$^{2}$\thanks{E-mail: song@uga.edu}  and
Simon Murphy$^{3}$
\\
$^{1}$School of Space Research, Kyung Hee University, 1732, Deogyeong-Daero, Giheung-gu, Yongin-shi, Gyunggi-do 17104, Korea\\
$^{2}$Department of Physics and Astronomy, The University of Georgia, Athens, GA 30602, USA\\
$^{3}$School of Science, University of New South Wales Canberra, ACT 2611, Australia
}

\date{Accepted XXX. Received YYY; in original form ZZZ}

\pubyear{2019}

\begin{document}
\label{firstpage}
\pagerange{\pageref{firstpage}--\pageref{lastpage}}
\maketitle

\begin{abstract}
We report the discovery of the oldest ($\sim$55 Myr) mid-M type star
known to host on-going accretion. 2MASS J15460752$-$6258042 (2M1546,
spectral type M5, 59.2\,pc) shows spectroscopic signs of accretion such as strong
\ha, \ion{He}{i}, and [\ion{O}{i}] emission lines, from which we
estimate an accretion rate of $\sim$10$^{-10}$ \msun\ yr$^{-1}$.
Considering the clearly detected infrared excess in all \emph{WISE}
bands, the shape of its spectral energy distribution and its age, we
believe the star is surrounded by a transitional disc, clearly with some
gas still present at inner radii. The position and kinematics of the
star from \emph{Gaia} DR2 and our own radial velocity measurements
suggest membership in the nearby $\sim$55 Myr-old Argus moving group.
At only 59pc from Earth, 2M1546 is one of the nearest accreting
mid-M dwarfs, making it an ideal target for studying the upper limit 
on the lifetimes of gas-rich discs around low mass stars.
\end{abstract}




\section{Introduction}

Stars inevitably harbour discs in the early stages of the star formation
process and the lifetime of such discs is one of the most important
parameters for understanding early stellar evolution. Moreover, because
gas giant planet formation must occur while the disc is gas-rich, the
disc lifetime is also a crucial parameter in planet formation scenarios.
By surveying young (2--30 Myr) clusters, \citet{hai01} showed the
lifetime of gas-rich discs is $\sim$2 Myr and ongoing accretion is rare
beyond ages of $\sim$10 Myr. \citet{mam09} noted that the disc
dissipation timescale should be dependent on the mass and hence spectral
type (temperature) of the host star, and \citet{rib15} showed that
high-mass ($>$2\,\msun) stars dispersed their discs up to twice as fast
as lower mass stars.

Recently, several examples of circumstellar disc accretion at ages
greater than $\sim$10 Myr have been identified (e.g., PDS 66- 20 Myr;
\citealt{mam02}, HD 21197- 30 Myr; \citealt{moo11}, 49 Ceti- 40 Myr;
\citealt{zuc12}, WISE J080822.18-644357.3- 45 Myr;
\citealt{silv16,mur18}, J0446A \& B, J0949A \& B; \citealt{silv20}), and
they have been treated as unusual anomalies without delving into the
problem of prolonged gas accretion at extreme ages. In spite of the
small number of cases, such old pre-MS stars hosting accretion discs can
be challenging to the hypothesis for the rapid planet formation
\citep{pfa19, man18, naj14, gre10} and provide the upper limit on the
life times of gas-rich discs.

The young M5 star 2MASS J15460752$-$6258042 (hereafter 2M1546) was
serendipitously found by us in a survey of low-mass moving group
candidate members from \emph{Gaia} DR2 (Lee et al. in preparation).
2M1546 was first observed spectroscopically and classified as a T Tauri
star with strong \ha\ emission by \cite{mis14} during their survey for
new symbiotic stars selected from the AAO/UKST SuperCOSMOS \ha\ Survey.
In this work, we analyse optical spectra of 2M1546, its IR excess and
spectral energy distribution, and evaluate its age based on the
kinematic membership of nearby young moving groups. From this analysis
we conclude that 2M1546 is the oldest (55 Myr)
and one of the nearest (59 pc) accreting M-type stars discovered to
date.

\begin{table*}
\centering
\caption{Summary of ANU 2.3-m/WiFeS observations of 2M1546.}
\label{tab:obs}
    \begin{threeparttable}
    \begin{tabular}{lccccccccc}
    \toprule
  UT date &      MJD & Grating & SpT & RV    & \vten[\ha] & \multicolumn{4}{c}{Equivalent width (\AA)\tnote{a}} \\ 
        \cmidrule(lr){7-10}
     &          &             &         & (\kms)  &  (\kms)         & \ha                   & \ion{He}{i} $\lambda$6678  & [\ion{O}{i}] $\lambda$6300    & \ion{Li}{i} $\lambda$6708 \\ \hline
2018 Jun 4 &   58273.5813  & R7000              & M5  & $-$4.4$\pm$0.9 &  366   & $-$210       & $-$1.6              &$-$3.6                      & 0.43 \\
2019 Feb 15 &  58529.6854 & R7000 \& B3000 & M5 & $-$3.5                & 310  & $-$120       & $-$0.8               & $-$3.8                   & 0.60 \\ \bottomrule 
       \end{tabular}
    \begin{tablenotes}
    \item[a] Negative values are in emission.
    \end{tablenotes}
    \end{threeparttable}
\end{table*}

\section{Spectroscopic observations and data analysis}

We observed 2M1546 using the Wide Field Spectrograph (WiFeS;
\citealt{dop07}) mounted on the ANU 2.3-m telescope at Siding Spring
Observatory during June 2018 and February 2019. A summary of these
observations is given in Table~\ref{tab:obs}. The B3000 and R7000
gratings provided wavelength coverage of 3500--6000~\AA\
($\lambda/\Delta\lambda\sim$3000) and 5300--7000~\AA\
($\lambda/\Delta\lambda\sim$7000), respectively.  
The raw data were reduced using the {\sc python}-based data reduction
pipeline, PyWiFeS  \citep{chi14}. The reduction process includes bad
pixel repair, bias, and dark current subtraction, flat fielding,
wavelength calibration, flux calibration, and data cube creation. The
wavelength calibration was performed using a series of Ne-Ar arc lamp
exposures, taken throughout the night. 

\begin{figure*}
\includegraphics[width=0.95\textwidth]{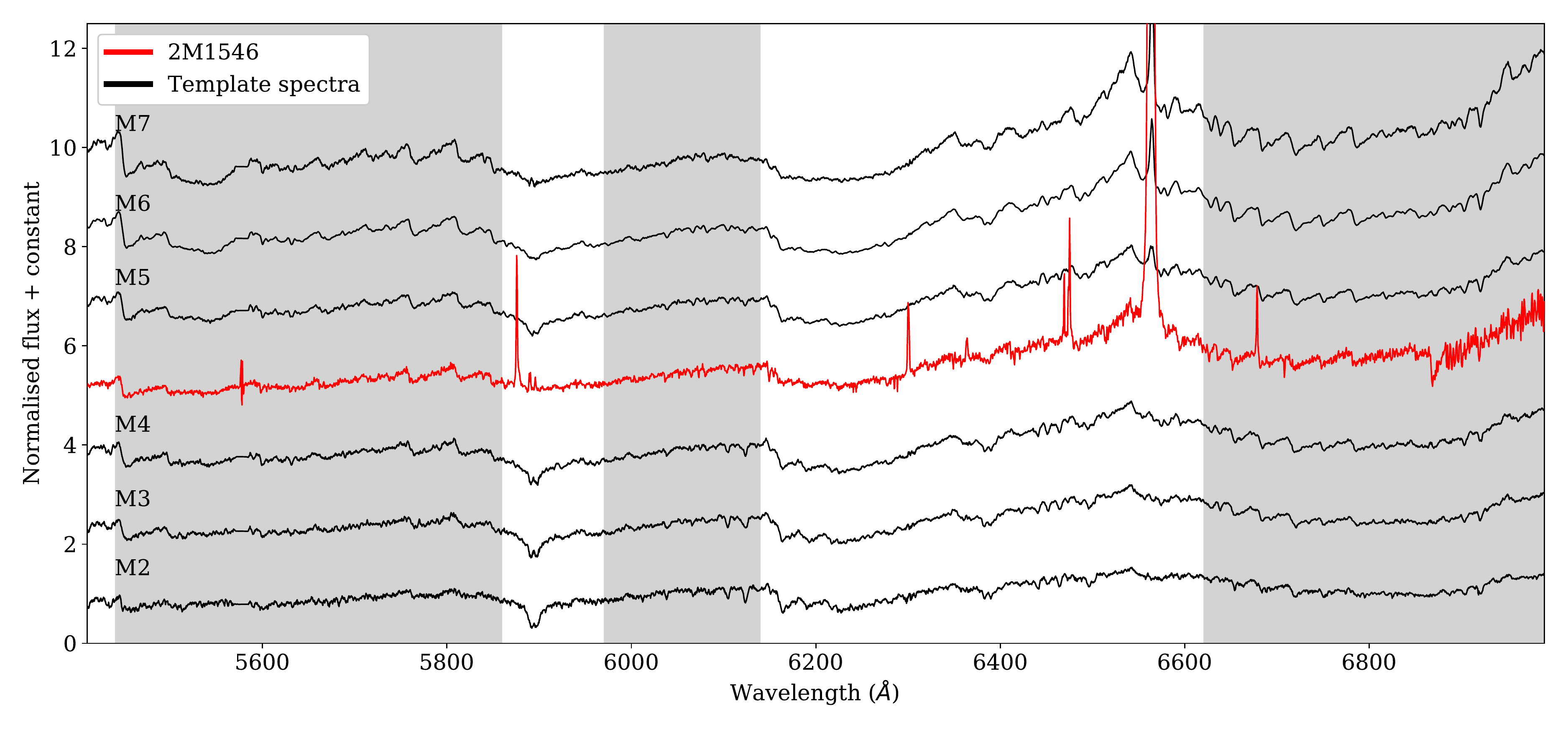}%
\caption{R7000 spectrum of 2M1546 (red) compared to the M-type templates
from \citet{boc07}. The templates are mean spectra of four thousand
dwarfs from the Sloan Digital Sky Survey (SDSS; \citealt{yor00}).  For
details of creating these template spectra, see \citealt{boc07}. The
spectrum of 2M1546 and the templates were normalised at 5800 \AA\ prior
to plotting.  In the spectral type determination, the overall shape of
the target and templates are compared focused more on the gray shaded area.}
\label{fig:r7000}
\end{figure*}

We compared the R7000 spectrum against the M-type spectral
templates of \citet{boc07} that were generated by utilizing more than
4,000 M-dwarfs spectra from the Sloan Digital Sky Survey (SDSS;
\citealt{yor00}).  Visual comparison of the 2M1546 spectra against a set
of M-dwarf template spectra shows the best match at M5.  2M1546 is
located well outside of any molecular cloud and, at a distance of only
60 pc, it should have negligible interstellar extinction.  Nevertheless,
the accretion causes ``veiling'', in which the accretion shock increases
flux in the UV/optical, while the surrounding accretion disc produces
additional flux to the photosphere in the IR \citep{vac11}. Veiling can
potentially affect our spectral type estimation because it will change
the spectral shape \citep{hh08,man13,ingleby14}. However, a change in
the spectral shape due to veiling occurs usually over a wide wavelength
range while our spectral type estimate is focused on rather narrow
wavelength ranges of 100 to 200\,\AA\ (e.g., shaded regions of
\ref{fig:r7000}).  Therefore, our estimate of M5 spectral type should
not be far from the truth and appropriate for the discussion in this
paper. A more robust, unambiguous spectral type estimate can be done in
the future by using a high spectral resolution echelle spectrum by
measuring line ratios of some temperature sensitive lines that are
adjacent in wavelengths so that veiling effect can be negligible.

The Li 6708\,\AA\ absorption feature was strongly detected in both epochs
of WiFES spectra (EW of 430 and 600\,m\AA\ ). Considering a typical equivalent
width measurement uncertainty ($\sim$50\,m\AA) in our WiFES spectra, the
observed difference is likely a real variability of the Li 6708 line strength
and related to a variable veiling. For an M5 star, the detected Li 6708
line strength indicates the undepleted level of Lithium in the atmosphere while
a "normal" 55\,Myr old M5 star should have depleted all Lithium already. An
on-going accretion must have replenished Lithium in the atmosphere so that the
6708 line was strongly detected in our observation.

\subsection{Emission Lines}

The magnetospheric accretion of the disc occurs
from the inner edge of the disc onto the surface of the central source.
This supersonic flow has nearly free-fall velocities, resulting in large
line widths of some emission lines.  In the region of the accretion
shock, on the other hand, some narrow emission lines are more likely
produced \citep{har16}.  The strength and/or shape of some lines such as
hydrogen recombination lines and \hei, Ca II, and Na I lines are known
as accretion tracers (see e.g., \citealt{muz98, ant11, bia14}).  The
observed spectra of 2M1546 cover \ha\ and \hei\ lines.

Both WiFeS observations show strong \ha\
emission, with equivalent widths in excess of 100\,\AA.  However,
because of the contrast effect caused by the depression of the stellar
continuum by molecular absorption, the equivalent widths of \ha\ lines
(EW(\ha)) in non-accreting stars are typically enhanced at later
spectral types \citep{whi03, mar98, bas95}.  \citet{bar03} presented an
EW(\ha) accretion criterion as a function of spectral type, where the
upper limit for accretors is EW(\ha) $\sim$ $-$20 \AA\ at M5. This was
corroborated by \citet{duc17} after an extensive analysis of known T
Tauri stars. With equivalent widths of $-$210\,\AA\ and $-$120\,\AA,
2M1546 clearly exceeds this criterion.

\begin{figure}
\includegraphics[width=0.9\columnwidth]{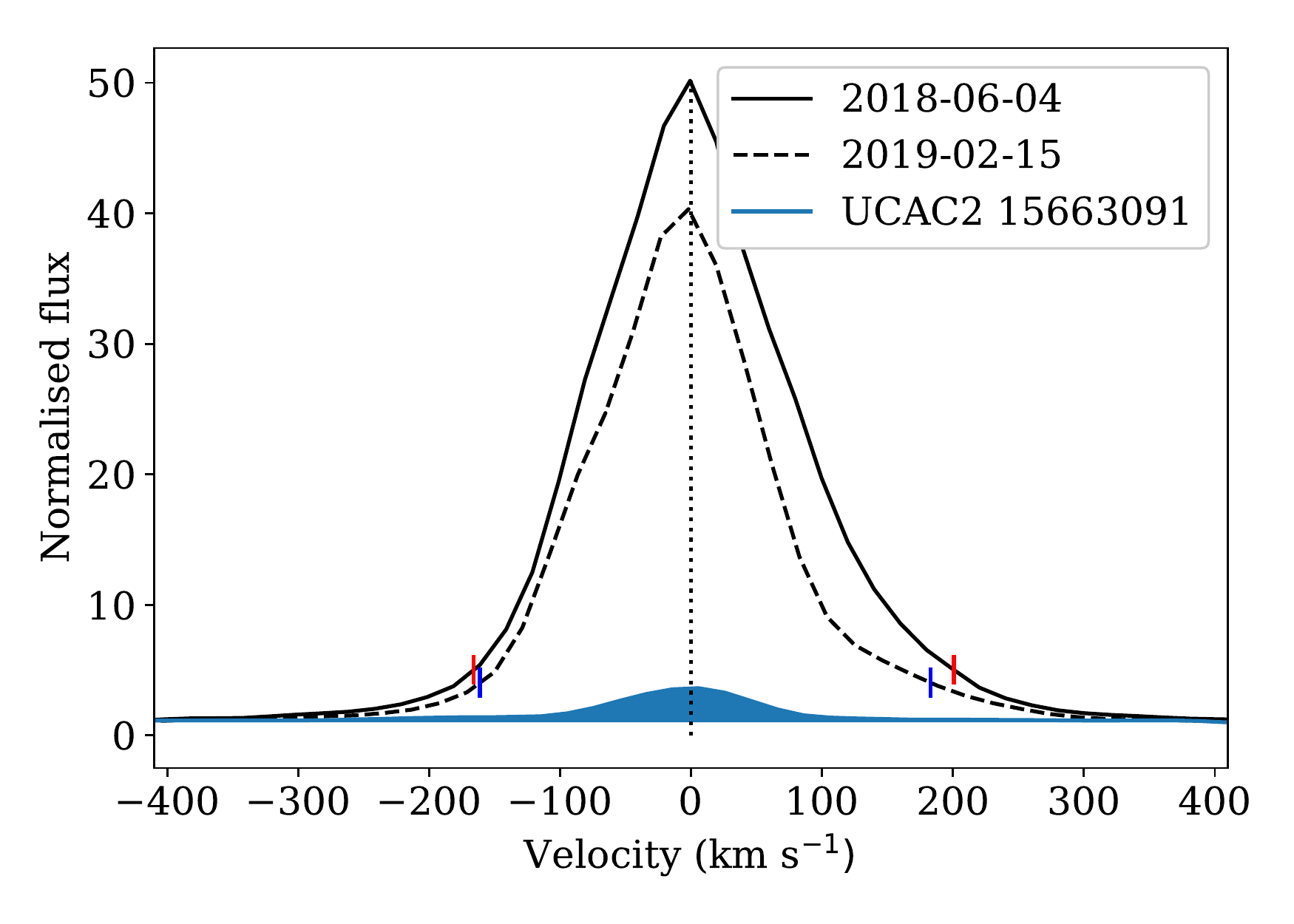}%
\caption{WiFeS/R7000 \ha\ velocity profiles of 2M1546.  Small vertical
lines (red and blue) indicate 10 per cent of the peak flux used for
measuring the \vten\ value.  The WiFeS line profile of the M5
non-accretor UCAC2 15663091 is also plotted for comparison.  All spectra
are shifted to the heliocentric rest frame.}
\label{fig:ha}
\end{figure}

\begin{figure*}
\includegraphics[width=0.8\textwidth]{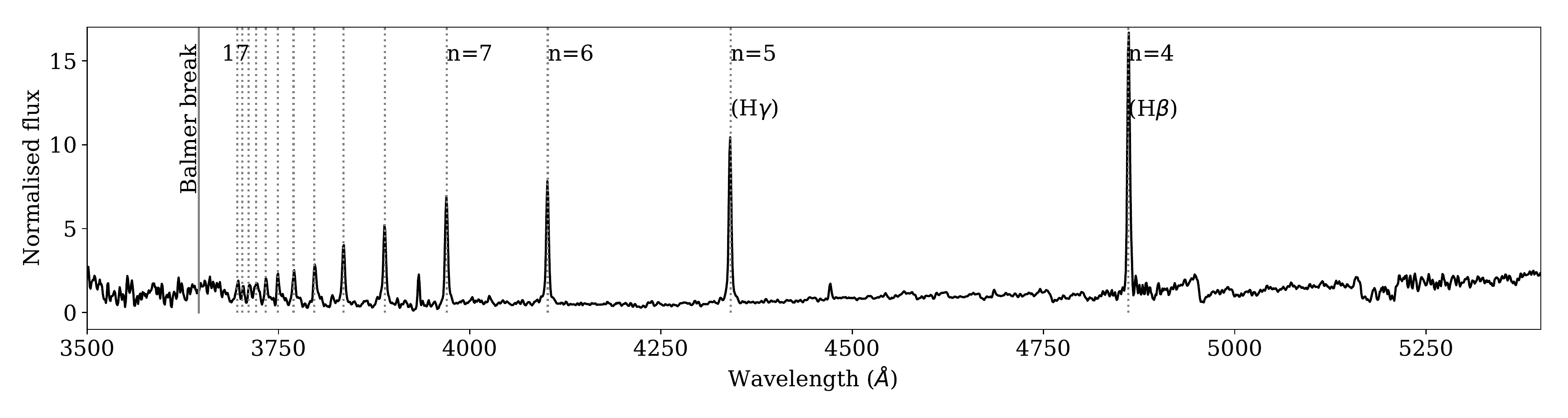}
\caption{The B3000 spectrum of 2MJ1546 showing Balmer series in emission up to $n=17$.}
\label{fig:b3000}
\end{figure*}

Although the contrast effect is considered, the evaluation of accretion
based on \ha\ strength has a caveat.  The chromospheric
activity--typically enhanced for young stars--generates a strong \ha\
emission similar to the case of accretion.  The line profile of \ha\ is
used for distinguishing between accreting and nonaccreting objects: disc
accretion generates broad and asymmetric \ha\ lines \citep{luh07}.
Fig.~\ref{fig:ha} shows the \ha\ velocity profiles for 2M1546, which are
asymmetric in both two observations.  The asymmetric feature can be
explained by the inclination effects and/or absorption by an
accompanying outflowing wind \citep{moh05, ale00}.  A quantitative
diagnostic value using \ha\ emission line profile is the full width of
\ha\ at 10 per cent of the line peak (\vten) \citep{whi03, moh05}.
However, the broadened \ha\ line profile is not always explained by the
accretion.  Fast rotator and binarity can broaden the \ha\ emission
line, which can be misdiagnosed as an accretor \citep{man13}.  In the
opposite way, the inclination of the accretion disc can produce a
narrower emission line along the line of sight, which can be
misdiagnosed as a nonaccretor \citep{moh05}.  Published \vten\ accretion
criteria vary from $200 < v_{10} <  270$ \kms\ \citep{fan13, jay03,
whi03}, independent of spectral type.  Measured \vten\ values for 2M1546 
are over 300 \kms (Table \ref{tab:obs}), which exceeds the strictest accretion criterion
(i.e., $>$270 \kms).  Using the \vten\ value and the accretion rate
(\macc) relation of \citet{nat04}, we estimate $\macc\ =
1.3\times10^{-10}$ \msun\ yr$^{-1}$.  Considering many M-type pre-MS
stars found to have accretion rates as low as $\sim$
2$\times10^{-12}$ \msun\ yr$^{-1}$
\citep{alc14, her09}, the rate we infer for
2M1546 suggests it is actively accreting from its inner disc. 

From our SED fitting, we obtain the bestfit stellar radius
(0.461\,$R_\odot$), effective temperature (2940 K), and luminosity
(0.014\,$L_\odot$). We alert readers that this SED fitting ignores any effects
from veiling and self-extinction so that they should be taken with caution.
Based on \citet{bar15}, these values are inconsistent with values from a
$\sim$55\,Myr old M5 star. These bestfit SED parameters are instead well
matched for parameters of a 10\,Myr old mid-M type star with mass of
0.1\,$M_\odot$. For this mass, a typical mass accretion rate is expected to be
$1.0\times10^{-10}$ \msun\ yr$^{-1}$ \citep{har16} which agrees well with our
obtained accretion rate.  As shown in the following section, 2M1546 is a highly
likely member of the 55\,Myr old Argus association. When a star has prolonged
accretion, then it may be conceivable that such a star has stellar parameters
of a younger star because of the accretion effect on its evolution.  One needs
higher spectal resolution data to confirm this possible effect of accretion by
independently obtaining a spectral type not affected by veiling.

\begin{figure*}
\includegraphics[width=0.95\textwidth]{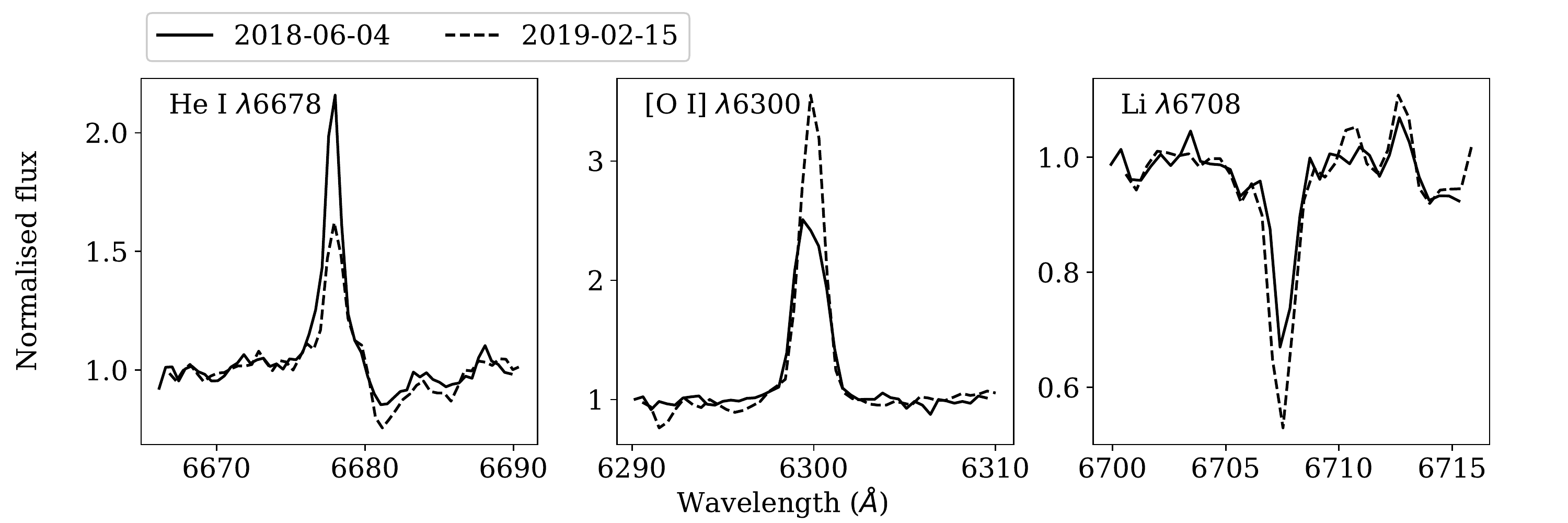}%
\caption{WiFeS/R7000 line profiles of \ion{He}{i} $\lambda$6678,
[\ion{O}{i}] $\lambda$6300, and \ion{Li}{i} $\lambda$6708 at the two observation
epochs.} \label{fig:otherlines}
\end{figure*}

Emission in \ion{He}{i} $\lambda$6678 is also reasonably well correlated
with accretion among very low mass stars and brown dwarfs \citep{moh05,alc14,alc17}.
\citet{giz02} showed that \ion{He}{i} $\lambda$6678 is generally present
in low levels (|EW(\ion{He}{i})|$\ll$1\,\AA) in older
chromospherically-active stars. 2M1546 shows strong \ion{He}{i}
$\lambda$6678 emission at both epochs (EWs of $-$0.8 \AA\ and $-$1.6
\AA), supporting the hypothesis of ongoing accretion (See
Table~\ref{tab:obs} and Fig.~\ref{fig:otherlines}). In
addition to strong \ha, we also detect other Balmer line emission up to
H17 (Fig.~\ref{fig:b3000}).

As shown in Figs.~\ref{fig:r7000} and \ref{fig:otherlines}, we detect
strong [\ion{O}{I}] emission at 6300~\AA\ at
both epochs.  Studies investigating the prominent
[\ion{O}{I}] $\lambda$6300 emission toward young stellar objects suggest
that jets, MHD winds, photoevaporative winds, and photodissociation of
OH in the disc surface layers are the possible origins of the line (e.g,
\citealt{ban19, nat14, rig13, ack05, sto00}).  \citet{sto00} predicts
that [\ion{O}{I}] $\lambda$6300 is dominated by OH photodissociation in the 
region with a line ratio [\ion{O}{I}] $\lambda$6300/[\ion{O}{I}] $\lambda$5577 
smaller than 10. Our spectra show a potentially weak ($\sim$1\,\AA\ ) detection 
of [\ion{O}{I}] $\lambda$5577 which results in the line ratio of about 3. This small
line ratio favours the other explanations for 
the origin of OI emission such as from winds and/or jets.

\begin{figure}
\includegraphics[width=\columnwidth]{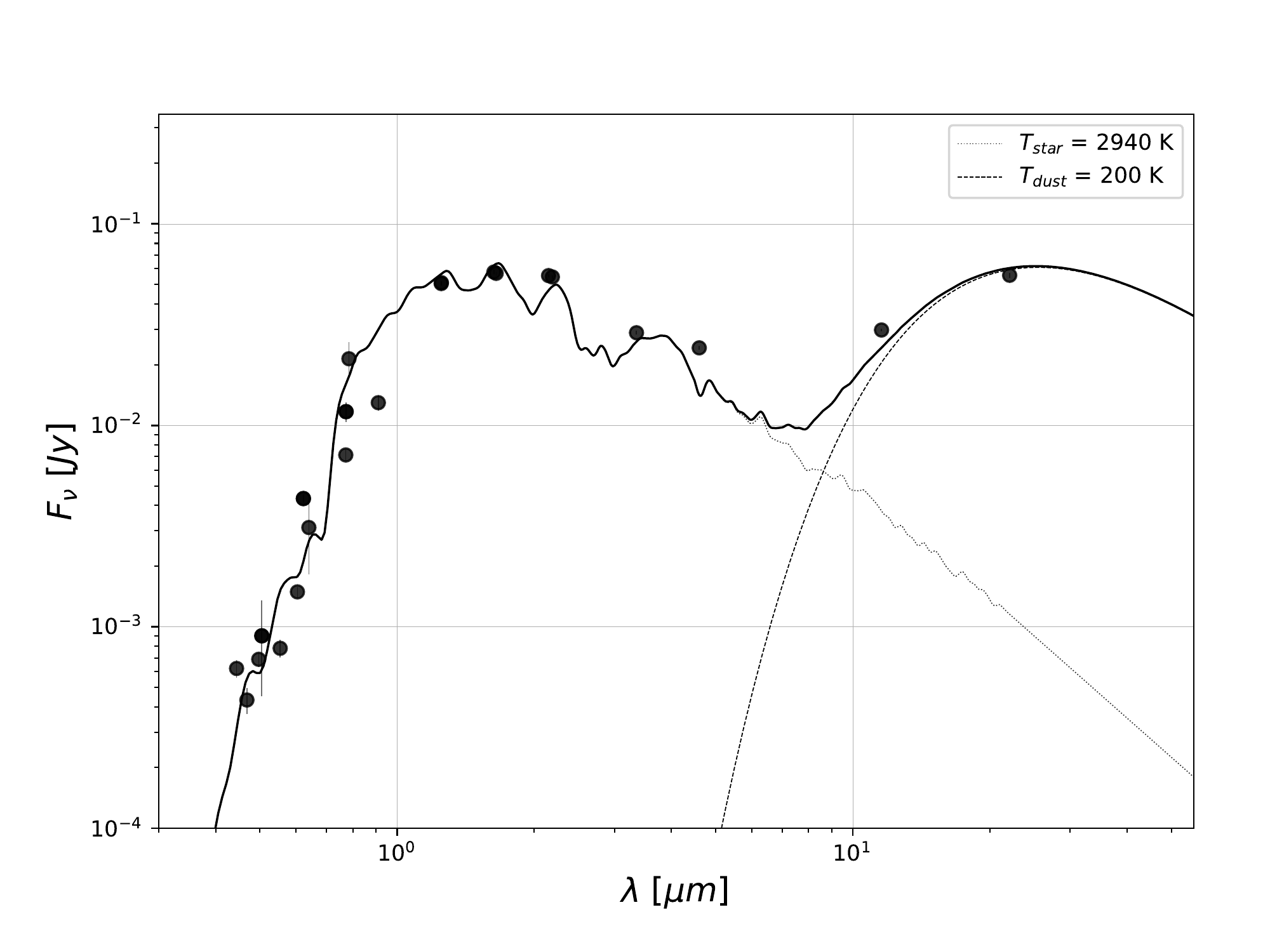}%
\caption{Spectral energy distribution for 2M1546 with the best-fitting
photosphere model of $T_{\rm star}$=2940 K (dotted line) and a 200 K
blackbody (dashed line) combined to produce a total model (solid line).}
\label{fig:sed}
\end{figure}

\section{Circumstellar disc emission}

Dust emission from the circumstellar disc is observed as the IR excess
of the spectral energy distribution (SED).  The SED of 2M1546 was
generated utilizing catalogue data from POSS-II ($JFN$;
\citealt{cab03}), SkyMapper DR1 ($griz$; \citealt{wol18}), {\it Gaia}
DR2 ($G\rm_{BP}$, $G\rm_{RP}$, $G$), 2MASS ($JHK_s$; \citealt{cut03}),
and All{\it WISE} ($W1-W4$; \citealt{cut14}). Examinations of high
angular resolution optical/near-IR images show no possible contaminating
source within the spatial resolution of WISE images. Therefore, we
conclude that there is no contamination and these photometric data are
only from 2M1546. We fit synthetic photometry derived from BT-Settl
(\citealt{all12}) using our SED fitting technique described in
\citet{rhe07}.  The resulting SED and the best-fitting model are
presented in Fig.~\ref{fig:sed}.

The best-fitting temperature ($T_{\rm star} = 2940$ K) is consistent
with the pre-MS temperature scale for young (10--30 Myr) stars derived
by \citet{pec13} for a spectral type of M5. We can satisfactorily fit
the \emph{WISE} $W3$ (12~\micron) and $W4$ (22~\micron) excesses with a
single blackbody of temperature $T_{\rm dust}\sim200$ K. With little
excess  emission at $\lambda<5$~\micron\ and a significant excess at
$\lambda>10$~\micron\ we classify 2M1546 as having a transitional disc
\citep{str89,wil11}, with clearly some material in the inner disc
remaining to drive the  accretion we observed.

\citet{mur18} investigated the disc surrounding the M5 Carina member
WISE J080822.18-644357.3 (hereafter WISE J0808),  which has a similar
SED and accretion characteristics to 2M1546.  They found a cold disc
component which was well fitted by a $\sim$240~K blackbody resulting
from populations of small dust grains released by sublimating
planetesimals. \citet{fla19} recently observed WISE J0808 with the
Atacama Large Millimetre/sub-millimetre Array (ALMA) but did not detect
any CO gas, indicating WISE J0808 may be at an evolutionary state
between a gas-rich transition disc and a gas-poor second-generation
debris disc. Considering the similarities between WISE J0808 and 2M1546,
the latter is also probably a borderline transition/debris disc
accreting the last of its inner gas on to the star.  More observations,
including from ALMA, are required before definitively assessing the
evolutionary stage of the disc around 2M1546.

\section{Age estimation based on moving group membership probability}

One of the key parameters for constraining disc models is stellar age
and, for young nearby stars within 150 pc, one of the most reliable
age-dating methods is to test whether the star is a member of one of
several young moving groups in the solar neighborhood with well
determined ages \citep[e.g.][]{zuc04,tor08}. Although the absolute age
scales from these moving groups should not be overinterpreted, relative
age ordering of several major moving groups is secure and can be used to
derive ages of member stars at mugh higher precision than possible for
an isolated star. The relative age rank for major moving groups are: TWA
$<$ BPMG $<$ Columba/Carina $<$ Argus $<$ AB\,Dor
\citep{zuc04,tor08,gag14,bel15}. We note that the existence of the Argus
group was challenged by \citet{bel15} and \citet{riedel17}, but
\citet{zuc19} reaffirm the Argus moving group. 

There are several moving group membership probability calculation tools
available in the literature (e.g. BANYAN II \& $\Sigma$ and BAMG 1 \& 2;
\citealt{gag14, gag18, lee18, lee19}).  The main differences between
these tools are their internal membership lists and how they calculate
the structural properties (mean $XYZ$ positions, $UVW$ velocities and
their distributions) of each moving group. BAMG~2 is the most recent
code for calculating kinematic memberships and was developed to be
self-consistent with respect to group memberships \citep{lee18}. For
this reason we adopt membership probabilities from it over those from
other tools.

\begin{figure*}
\includegraphics[width=0.95\textwidth]{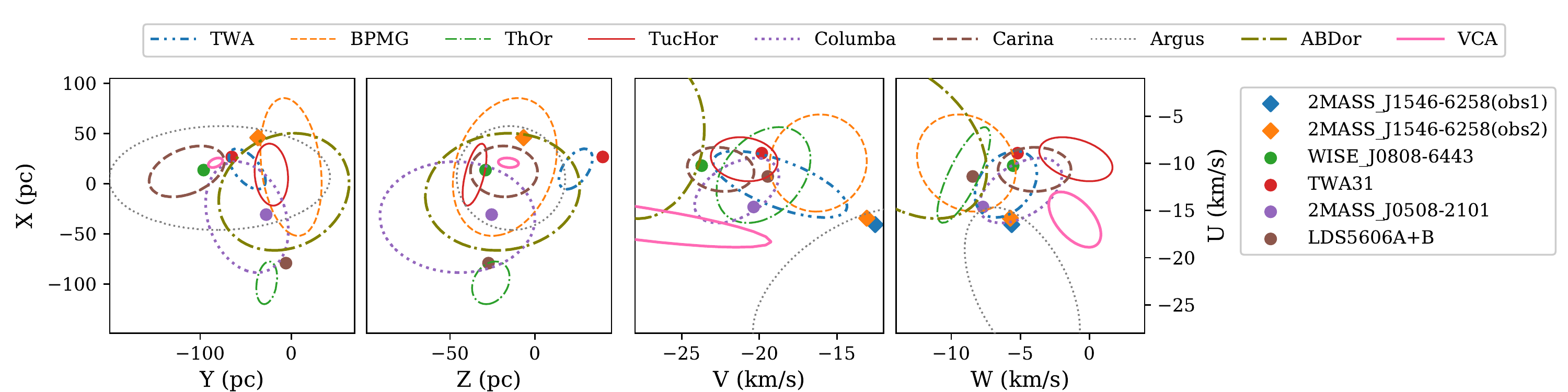}
\caption{Accreting mid-M type stars and models of NYMGs in the $XYZUVW$ spaces. }
\label{fig:xyzuvw}
\end{figure*}

Table~\ref{tab:prob} presents membership probabilities using our two
measured RV values.  BAMG 2 evaluates 2M1546 as an Argus member while
the two BANYAN models provide a very small Argus membership probability,
with the star most likely being a field object.  If 2M1546 was an old
field star, then its ongoing accretion is even more interesting yet more
challenging to understand. If an isolated accreting pre-MS star, then
2M1546 belongs to a rare group of nearby ($\lesssim$60\,pc) youngest stars such as some 
accreting TWA members. Fig.~\ref{fig:xyzuvw} compares the
heliocentric position ($XYZ$) and velocity ($UVW$) of 2M1546 to the
moving group models used by BAMG 2. The position and velocity of 2M1546
are well matched to the $\sim$55\,Myr-old Argus
group, as expected from the membership probabilities in
Table~\ref{tab:prob}.

\begin{table*}
\caption{Kinematic membership probabilities.}
\label{tab:prob}
\begin{threeparttable}
\setlength{\tabcolsep}{3pt}
\begin{tabular}{p{0.17\linewidth}p{0.32\linewidth}p{0.42\linewidth}}
\toprule
& \multicolumn{2}{c}{Membership probability: Group (per cent)} \\
\cmidrule(lr){2-3} 
RV (\kms) & $-4.4\pm0.9$ & $-3.5\pm1$ \\ 
BAMG 2 & Argus(95) + Field(5) & Argus(83) + BPMG(14) \\
BANYAN II & Field(78) + Argus(22) & Field(89) + Argus(9)  \\
BANYAN $\Sigma$ & Field(94) + Argus(2) & Field(87) + UCL(8) + Argus(5)  \\
\bottomrule
\end{tabular}
\begin{tablenotes}
\item \emph{Note.} Since only groups with membership probabilities $>$2
per cent are presented here, the sum of probabilities is smaller than
100 per cent.  
\end{tablenotes}
\end{threeparttable}
\end{table*}

\section{Discussion and Conclusions}

\begin{table}
\caption{Summary of 2M1546 parameters.}
\label{tab:sum}
\begin{threeparttable}
    \begin{tabular}{lcc}
    \toprule
    Parameter & Value & Units \\ \hline
    R. A. & 15:46:07.52\tnote{a} & hh:mm:ss.ss \\
    Dec. & $-$62:58:04.2\tnote{a}  & dd:mm:ss.s \\
    SpT & M5 \\
    $\mu_{\alpha}\rm cos\delta$ & $-42.7\pm0.1$\tnote{a} & mas yr$^{-1}$ \\
    $\mu_{\delta}$ & $-61.5\pm0.1$\tnote{a} & mas yr$^{-1}$ \\ 
    Distance & $59.2\pm0.3$ \tnote{a} & pc \\
    RV & $-4.4$, $-3.5$ & \kms \\
    $T_{\rm eff}$ & 2940 & K \\
    log($L/L_{\odot}$) & $-$1.83 & dex \\
    \macc & 1.31$^{-10}$ & \msun\ yr$^{-1}$ \\
    Mass & 0.11 & \msun \\
    Age & 55 & Myr \\
    $X, Y, Z$ & 45.9, $-$36.7, $-$6.7 & pc \\
    $U, V, W$ & $-$16.5, $-$12.5, $-$5.6\tnote{b}& \kms \\
                 & $-$15.8, $-$13.1, $-$5.7\tnote{c} & \kms \\              
    \bottomrule
\end{tabular}
\begin{tablenotes}
    \item[a] \citet{gai18}
    \item[b] Using $\textrm{RV} = -4.4$ \kms 
    \item[c] Using $\textrm{RV} =  -3.5$ \kms 
\end{tablenotes}

\end{threeparttable}
\end{table}

We summarize the important properties of 2M1546 from the literature and
this work in Table~\ref{tab:sum}. The proximity (59~pc), late spectral
type (M5),  old age (55\,Myr), mid-IR excess
emission, and ample evidence of on-going accretion observed in 2M1546
make it an interesting laboratory for studying prolonged disc accretion.
It is the oldest known accreting star to date,  and can therefore be
used for studying the upper limit on the lifetimes of gas-rich discs.
The kinematic properties of the star are consistent with the Argus
moving group and its estimated age of $\sim$55\,Myr.

Over the past several years, several mid-M type accretors at various
ages have been reported in the literature (Table~\ref{tab:othersrcs}).
All of the stars in Table~\ref{tab:othersrcs} show strong mid-IR excess
emission at $W3$ and $W4$ and signs of ongoing accretion.  Among these
sources, WISE J0808 appears to be the most similar to 2M1546, with
\vten\ velocity widths in excess of 300~\kms. 2MASS J1239$-$5702 and
2MASS J1422$-$3623 have slightly weaker \ha\ equivalent widths and
\vten\ values comparble to 2M1546. However, they are believed to be
younger (10$-$17 Myr) members of LCC or UCL subgroups of the Sco-Cen OB
association \citep{mur15}. While LDS 5606A+B are also actively accreting
sources, they are binaries and hence their accretion might have been
affected by their companions.  2MASS J1337$-$4736 is claimed to be
accreting based on its asymmetric \ha\ profile and \emph{WISE} excess
\citep{rod11, sch12b}. \citet{zuc15} found that this star has a distant
companion. While not having a measured \vten\ value, TWA 31 has a strong
EW(\ha) value and significant excess at $W3$ and $W4$, supporting an
accretion hypothesis \citep{sch12a, zuc15}. The remaining three stars
seem to show marginal signs of accretion. 2MASS J0844$-$7833 and 2MASS
J0508$-$2101 barely exceed the lower limit for accretion criterion in
terms of \vten\ and/or EW(\ha).  Compared to other accretion sources,
2MASS J0501$-$4337 also has a significantly lower EW(\ha) and a \vten\
width that barely exceeds the accretion criterion of \citet{bar03}.

With the exception of WISE J0808, the inferred ages for these stars are
all smaller than the age of the $\beta$ Pic Moving Group ($\lesssim25$ Myr).
Therefore, the slightly prolonged accretion at these stars could have
been accepted as an unusual phenomenon seen in some rare outliers. Now,
we see at least a handful of adolescent mid-M type stars with clear
signs of ongoing accretion, suggesting that accretion can be maintained
for several tens of million years around low-mass stars under certain conditions. The
discovery of these older M-type accretors could eventually reveal a less
efficient process of removing circumstellar gas and dust around low-mass
stars. We note that the plethora of M5 accretors at ages of
30--55\,Myr coincides with the mass boundary at
which stars become fully convective in their interiors. 

\begin{table*}
\centering
\caption{Accreting mid-M type stars from the literature.}
\label{tab:othersrcs}
\begin{threeparttable}
    \begin{tabular}{lcccccccc}
    \toprule
    Name                                & SpT  & Dist.        & RV             & EW(\ha)       & \vten[\ha]    & Age from Lit.     &   Age from this work\tnote{a} & Refs.\\ 
                                        &      & (pc)         & (\kms)         & (\AA)         & (\kms)        & (Myr; group)      & (Myr; group)                           \\     \hline
2MASS J08440915$-$7833457               & M4.5  & 98          & $-$            &  60           &      212      & 8--14 ($\eta$Cha) & $-$\tnote{b}  & 1 \\
    TWA 31                              & M4.2  & 81          &10.5$\pm$0.4    & 115           &    $-$        & 7--13 (TWA)       & 7--13 (TWA)  & 2 \\
  2MASS J13373839$-$4736297             & M3.5 & 126          &      $-$       & 13.7          & $-$           & 10--17 (LCC)      & $-$\tnote{b}         & 3, 4\\
  2MASS J12392312$-$5702400             & M5   & 180          & 16$\pm$2       & 27--63        & 238$-$331     & 10--17 (LCC)      & $-$\tnote{b}         & 5\\
  2MASS J14224891$-$3623009             & M5   & $-$          & 9$\pm$2        & 33--91        & 236$-$341     & 10--16 (UCL)      & $-$\tnote{b}           & 5 \\ 
 2MASS J05082729$-$2101444              & M5    & 49          & 24.9$\pm$0.9   & 25            &    197        & 12--25 (BPMG)     & 30--44 (Columba) & 1  \\
    LDS 5606 A                          & M5    & 84          &14.9$\pm$0.8    & 99--135       &    250        & 12--25 (BPMG)     & 30--44 (Columba) & 6, 7\\
    LDS 5606 B                          & M5    & 84          &14.9$\pm$0.4    & 25            &      135      & 12--25 (BPMG)     & 30--44 (Columba) & 6, 7\\
 2MASS J05010082$-$4337102              & M4.5  & $-$         & $-$            &  7.6          &      210      & 30--44 (Columba)  & $-$\tnote{c}    & 8 \\   
       WISE J080822.18$-$644357.3       & M5    & 90\tnote{d} & 22.7$\pm$0.5   & 65--125       &  300--350     & 30--49 (Carina)   & 30--49 (Carina) & 9 \\
 2MASS J15460752$-$6258042              & M5    & 59          & $-$4.4, $-$3.5 &  120--210     & 310--366      &  $-$              &  40--69 (Argus) & 10 \\      
      J0446A                            & M6    & 83          &26.7$\pm$16.8   & 10.4          & 210           & 30--44 (Columba)  & $-$          & 11 \\
      J0446B                            & M6    & 82          &29.8$\pm$16.8   & 16.8          & 239           & 30--44 (Columba)  & $-$          & 11 \\
      J0949A                            & M4    & 79          &22.4$\pm$16.7   & 110           & 367           & 30--49 (Carina)   & $-$          & 11 \\
      J0949B                            & M5    & 78          &20.5$\pm$16.8   &  24           & 305           & 30--49 (Carina)   & $-$          & 11 \\
    \bottomrule
    \end{tabular}
\begin{tablenotes}
\item[a] Ages are obtained based on the moving group membership probabilities calculated using BAMG 2 except where marked. 
\item[b] $\eta$ Cha, LCC, and UCL are not included in BAMG 2. BANYAN $\Sigma$ suggests that 2MASS J0844-7833, J1337-4736, J1239-5702, and J1422-3623 are likely members of $\eta$ Cha  ($P=62$ per cent), LCC  ($P=63$ per cent), LCC ($P=52$ per cent), and UCL ($P=84$ per cent), respectively.
\item[c] Since a \emph{Gaia} proper motion for this star not exist, the Bayesian membership probability cannot be calculated. 
\item[d] Kinematic distance based on membership in Carina \citep{mur18}. \\
{\bf Notes.} References for group age: $\eta$ Cha- \citealt{bel15}; TWA-\citealt{bel15}; LCC- \citealt{son12}, \citealt{pec16}; UCL- \citealt{son12}, \citealt{pec16}; BPMG- \citealt{son03}, \citealt{bel15}; Columba- \citealt{tor08}, \citealt{bel15}; Carina- \citealt{tor08}, \citealt{bel15}; Argus- \citealt{bel15}, \citealt{zuc19} \\
{\bf References.} (1) \citealt{sch19}; (2) \citealt{sch12a}; (3) \citealt{sch12b}; (4) \citealt{rod11}; (5) \citealt{mur15}; (6) \citealt{rod14}; (7) \citealt{zuc14}; (8) \citealt{bou16}; (9) \citealt{mur18}; (10) This work.; (11) \citealt{silv20}
\end{tablenotes}
\end{threeparttable}
\end{table*}

\section*{Acknowledgements}

We thank Michael Bessell for obtaining the second epoch B3000 and R7000 spectra of 2M1546 in 2019 February.
This work was supported by the Basic Science Research Program through the National Research Foundation of Korea (grant No. NRF-2018R1A2B6003423).








%
%


\bsp	
\label{lastpage}
\end{document}